\documentstyle[preprint,aps,eqsecnum]{revtex}

\newcommand{\lbar}{\overline}

\newcommand{\be}{\begin{equation}}
\newcommand{\ee}{\end{equation}}
\newcommand{\bea}{\begin{eqnarray}}
\newcommand{\eea}{\end{eqnarray}}

\newcommand{\bhz}{\hat{\mbox{\boldmath $z$}}}

\newcommand{\bp}{{\mbox{\boldmath $p$}}}
\newcommand{\bk}{{\mbox{\boldmath $k$}}}

\newcommand{\blambda}{\mbox{\boldmath $\lambda$}}

\newcommand{\sblambda}{\mbox{\boldmath \scriptsize$\lambda$}}

\def\lsim{\ \rlap{\raise 3pt \hbox{$<$}}{\lower 3pt
\hbox{$\sim$}}\ }
\def\gsim{\ \rlap{\raise 3pt \hbox{$>$}}{\lower 3pt
\hbox{$\sim$}}\ }
\def\vev#1{\langle #1 \rangle}

\def\sslash{s\!\!\!\slash}
\def\kslash{k\!\!\!\slash}

\def\pslash{p\!\!\!\slash}
\def\delslash{\partial\!\!\!\slash}

\def\half{{\textstyle{1 \over 2}}}
\def\ihalf{{\textstyle{i \over 2}}}
\def\eighth{{\textstyle{1 \over 8}}}  

\def\npb#1{Nucl.\ Phys.\ {\bf B#1}}
\def\plb#1{Phys.\ Lett.\ {\bf B#1}}
\def\prd#1{Phys.\ Rev.\ {\bf D#1}}
\def\prl#1{Phys.\ Rev.\ Lett. {\bf#1}}

\def\sjnp#1{Sov. J. Nucl. Phys. {\bf #1}}

\begin{document}

\draft
                          
{\tighten
\preprint{\vbox{
                \hbox{UdeA-PE-00/001}
                \hbox{hep-ph/0002026}
                \hbox{January, 2000}}}

\title{Effects of a general set of interactions \\ 
on neutrino propagation in 
matter\footnote{\footnotesize 
Based on the article [1] written in collaboration
with Sven Bergmann and Yuval Grossman.}}

\author{Enrico Nardi}

\address{Departamento de F\'\i sica \\
Universidad de Antioquia,
A.A. {\it 1226}, \ Medell\'\i n,  \ Colombia \\ 
E-mail: enardi@naima.udea.edu.co}


\maketitle

\vskip-.3truecm
\begin{abstract}%
An  analysis of the effective potential for 
neutrino propagation in matter, assuming a generic set of 
Lorentz invariant non-derivative interactions is presented. 
In addition to vector and axial vector couplings, also tensor  
interactions can give coherent effects if the medium is polarized,    
and the components of a tensor potential transverse to the  
direction of neutrino propagation can induce a neutrino spin-flip.
\end{abstract}
} 

\section{Introduction}

Neutrino physics currently provides the strongest experimental
evidence for physics beyond the Standard Model (SM).
The atmospheric neutrino anomaly~\cite{Atmospheric}    
and the solar neutrino problem~\cite{SN} 
are best explained by neutrino oscillations. This 
require massive neutrinos that mix, and hence 
physics beyond the SM.  
When neutrinos propagate in matter, the physics of neutrino oscillations 
can be  very different from the case of vacuum propagation.
This is because coherent interactions with the background give to
the neutrino an ``index of refraction" that depends on the 
type of background and on the neutrino flavor. 
For example, in  normal matter only electron 
neutrinos have SM charged current interactions, and thus  
the effective $\nu_e$ mass is enhanced with respect to the other 
flavors. 
This allows for the possibility of level crossing between 
different neutrino eigenstates in matter, 
and can result in a significant
amplification of neutrino oscillations.
This is known as the MSW effect.~\cite{MSW}  
For light sterile neutrinos also neutral current interactions 
are important.~\cite{sterile} Finally, in a
polarized medium the neutrino effective mass depends also on the
average polarization of the background and on the angle between the
neutrino momentum and the polarization
vector.~\cite{Nunokawa}

New physics models that imply massive neutrinos  
often predict also new neutrino
interactions, that can significantly modify the SM 
picture.~\cite{NP-MSW,BergmannKagan}
For example, non-universal interactions may give rise to matter
effects that distinguish between muon and tau neutrinos.  Lepton
flavor violating interactions can induce an effective mixing in
matter, allowing for a resonant conversion even in the absence of
vacuum mixing.  The two effects combined together can induce
neutrino flavor transitions even for massless neutrinos.
Most of the discussions of these non standard effects 
assume new interactions just of vector and axial vector types.  
However, recently the possible effects of a much more 
general set of interactions have been analyzed.~\cite{BGN}
In this talk we discuss the main results of 
this investigation.


\section{Neutrino propagation in matter with general interactions}

Our aim is to study neutrino propagation in matter
in the presence of the most general pointlike and Lorentz invariant
four-fermion interaction with the background fermions ($f=e, p, n,
\nu$). We assume an interaction Hamiltonian of the form 
\be \label{Hint}
{\cal H}_{\rm int} = 
{G_F \over\sqrt{2}} 
\sum_a (\bar \nu\,\Gamma^a\,\nu)\, 
\left[\bar \psi_f\,\Gamma_a\, (g_a + g'_a
\gamma^5)\,\psi_f\right]\, + {\rm h.c.}\,, 
\ee
where
$\Gamma^a=\{I,\gamma^5,\gamma^\mu,\gamma^\mu\gamma^5,\sigma^{\mu\nu}\}$,
$\sigma^{\mu\nu}=\ihalf[\gamma^\mu,\gamma^\nu]$ and  
$a=\{S,P,V,A,T\}$. The Fermi constant $G_F$ has been
factored out so that all the couplings are dimensionless.  
In general,
$\nu$ is a vector of the different neutrino types, and $g_a, g'_a$ are
10 matrices in the space of neutrino flavors that describe the
coupling strengths.  Note that new interactions can include
both flavor diagonal and off-diagonal couplings.  
To derive the equation of motion 
for the neutrino propagation in matter we first average the
effective interactions over the background fermions.  
We select only coherent transitions, which leave the
many-fermion background system in the same state, 
since incoherent effects become negligible after
averaging. In particular, while we allow for neutrino 
spin-flips, we require that the background 
fermions  do not change their spin. 
Accordingly, we introduce    
matrix elements of the fermion currents between 
initial and  final states with the same quantum numbers   
\be \label{matrixelement}
{\cal M}_a^f \equiv 
\langle f,\bp,\blambda|\bar \psi_f\,\Gamma_a\,(g_a + g'_a
\gamma^5)\,\psi_f|f,\bp,\blambda 
\rangle \,,   
\ee
where $\bp$ and $\blambda$ denote respectively the momentum 
and polarization vectors of the background fermion $f$. 
The expectation value of ${\cal M}_a^f$, averaged over 
the fermion distribution
$\rho_f(\bp,\blambda)$ reads  
\be \label{Va}
V_a^f = {G_F \over \sqrt{2}} 
\sum_{\sblambda} \int {d^3 p \over (2\pi)^3} 
\rho_f(\bp,\blambda)
{\cal M}_a^f \,. 
\ee
The effect of the medium on the neutrino propagation in 
the presence of the general interactions~(\ref{Hint}) 
is then described  by the interaction Lagrangian
\be \label{Lint} 
-{\cal L}_{int} = \sum_{a,f} 
(\bar \nu\,\Gamma^a\,\nu)\, V^f_a \,. 
\ee
The computation of the various ${\cal M}_a^f$ is
straightforward.~\cite{BGN} 
After performing the contractions $\Gamma^a \, V_a^f$
in~(\ref{Lint}) we obtain  
\bea 
\Sigma^{SP} &\equiv& \  \left[V^S + V^P \, \gamma^5 \right]
= 
{G_F \over \sqrt{2}} \, n_f 
\left<{m_f \over E_f} \right> \, \left(g_S + g'_P \, \gamma^5 \right) 
\label{VSP}  \\
\Sigma^{VA} &\equiv& \gamma^\mu\left[V_\mu^V+V_\mu^A \,\gamma^5\right]
\nonumber \\ 
&=&  
{G_F \over \sqrt{2}} \, n_f \left[
\left<{\pslash\over E_f}\right> \, \left(g_V +g'_A\,\gamma^5
\right)
%
%
~+ m_f \, \left< {\sslash \over E_f}\right> 
\left(g'_V + g_A \, \gamma^5 \right) \right] 
\label{VVA}  \\
\Sigma^T&\equiv& \varsigma^i\left[V^B_i + i V^E_i \, \gamma^5
\right]
= {G_F \over \sqrt{2}} \, n_f 
\left<{[\sslash, \pslash]\over E_f} \right> \, 
\left(g'_T + g_T \, \gamma^5 \right)\,, 
\label{VTT}  
\eea
where $\varsigma^i \equiv \mbox{diag} \, (\sigma^i, \sigma^i)$. 
The spin vector 
$s\,$  satisfies $s^2=-1\,$ 
and $s_\mu \, p^\mu=0\,$   
(the explicit expression can be found  in~\cite{BGN}).  
We have also introduced 
\be
n_f = \sum_{\sblambda} \int {d^3 p \over (2\pi)^3} 
\rho_f(\bp,\blambda) \, , 
\qquad
\vev{x}={1 \over n_f}\sum_{\sblambda} \int {d^3 p \over (2\pi)^3} 
\rho_f(\bp,\blambda) \, x(\bp,\blambda) \,   
\ee
to denote, respectively, the number density $n_f$ of the 
fermion $f$ and the
average of some function $x(\bp,\blambda)$ over the 
fermion distribution. 
In~(\ref{VTT})  we have decomposed
the tensor term $V^T_{\mu\nu}$ in analogy to the electro-magnetic
field tensor $F_{\mu\nu}$, as $V^B_i = \epsilon_{ijk} \, V^T_{jk}$ and
$V^E_i = 2\,V^T_{0i} $.  Note that the second equality in~(\ref{VTT})
makes apparent that the tensor interaction can contribute only in the
presence of a polarized background.

%
The equation of motion 
for the neutrino propagation can be deduced from the Lagrangian
\be \label{Lagr}
{\cal L} = {\cal L}_{free} + {\cal L}_{int}=
\bar{\nu} (i \delslash - m - \Sigma) \nu 
\ee
where  the matrix of the potentials 
$
\Sigma  \equiv \Sigma^{SP} + \Sigma^{VA} +  \Sigma^T\, 
$
depends on the background density and polarization, and in general
will vary along the neutrino propagation path.  In the general case
both $\Sigma$ and $m$ are matrices in the space of neutrino types.
In the chiral basis the interaction part in~(\ref{Lagr}) reads 
\be \label{Lint2}
-{\cal L}_{int} = \bar \nu \, \Sigma \, \nu =
\pmatrix{\nu^\dagger_L \cr \nu^\dagger_R}^T
\pmatrix{V^{LL}_\mu \bar\sigma^\mu &  V^{LR}_\mu \sigma^\mu \cr
         V^{RL}_\mu     \sigma^\mu &  V^{RR}_\mu \sigma^\mu \cr}
\pmatrix{\nu_L \cr \nu_R},
\ee
where $\sigma^\mu\, (\bar\sigma^\mu)=(\sigma^0\,, (-)\, \sigma^i)$ 
with $\sigma^0=I$,  and 
\bea
V^{LL}_\mu &\equiv& V_\mu^V - V_\mu^A\,,~~~~~
V^{RL}_0 \equiv V^S - V^P\,,~~~~~ 
V^{RL}_i \equiv V_i^B-i\,V_i^E\,,
\\
V^{RR}_\mu &\equiv& V_\mu^V + V_\mu^A\,, ~~~~~ 
V^{LR}_0 \equiv V^S + V^P\,, \label{LRSP} ~~~~~
V^{LR}_i \equiv V_i^B+i\,V_i^E\,.  \label{LRT} 
\eea
It is apparent that the (axial)vector
potentials appearing in the diagonal entries in (\ref{Lint2}) 
couple neutrinos of the same chirality, 
while the (off-diagonal) (pseudo)scalar and tensor potentials 
couple neutrinos of opposite chirality.
The equations of motion for neutrinos and antineutrinos 
derived from~(\ref{Lagr}) read 
\be \label{EOMnu}
\gamma_0 (\kslash - m - \Sigma) u = 0\,, \qquad\quad 
\gamma_0 (\kslash + m + \Sigma) v = 0\,.   
\ee
Note that the signs of $m$ and $\Sigma$ are opposite for the
antineutrinos. The dispersion relations for the neutrino propagation
are given by the solutions of
\be \label{Disp} 
{\rm det}\,[{\cal O}] = {\rm det}\,[\gamma_0(\kslash - m -\Sigma)] = 0. 
\ee 
Let us take the neutrino momentum $\bk = k \bhz\,$  
along the $z$-axis. 
Assuming that  $V^{V,A,T}, m \ll
E\,$ and neglecting terms of ${\cal O}(m/E)$ 
the relevant terms    
in~(\ref{Lint2}) are $V^{LL}_{0,3}\,$, $V^{RR}_{0,3}$ and the tensor
components $V^{LR}_{1,2}\,$,  $V^{RL}_{1,2}$ 
transverse with respect to the neutrino propagation direction. 
Solving the determinant equation (\ref{Disp}) yields the neutrino
energies 
\be \label{Solution}
E_\pm  =
k + {m^2\over 2 k}+ \half \left[V^{LL}_{0-3} + V^{RR}_{0-3} \pm 
\sqrt{\left(V^{LL}_{0-3} - V^{RR}_{0-3}\right)^2 + 
      4\, V^{LR}_+ \, V^{RL}_-} \, \right] \,,   
\ee
where $V_{0 \pm 3} \equiv V_0 \pm V_3\,$ and $V_\pm \equiv V_1 \pm i
V_2$. In (\ref{Solution}) 
the plus (minus) sign refers to neutrinos that are mainly
left(right)-handed states.  Eliminating the two helicity suppressed
states from the equations of motion we obtain a Schr\"odinger-like
equation that governs the neutrino propagation:
\be \label{EOMLR}
i {d \over dt} \pmatrix{\nu_L \cr \nu_R} = 
{\cal H}_\nu \pmatrix{\nu_L \cr \nu_R} ~~~~
\mbox{with}~~~~{\cal H}_\nu =
k + {m^2 \over 2k} +
\pmatrix{V^{LL}_{0-3} & V^{LR}_+ \cr 
         V^{RL}_-     & V^{RR}_{0-3} \cr} \,.
\ee
The two  eigenvalues of the effective Hamiltonian ${\cal H}_\nu$ are
the solutions~(\ref{Solution}) of the determinant equation~(\ref{Disp}).
The results for the  antineutrinos
can be obtained from~(\ref{EOMLR}) and~(\ref{Solution}) by changing the
sign of the potentials ($V \to -V$). 
In the case of more than one neutrino
flavor~(\ref{Solution}) is a matrix equation in the space of the
neutrino types. 
In the one flavor case, the energy gap between the two states
is
\be \label{DeltaE}
\Delta E_\nu =
\sqrt{\left(V^{LL}_{0 - 3} - 
V^{RR}_{0 - 3} \right)^2 + 4\, V^{LR}_+ \, V^{RL}_-} \, . 
\ee
In the limit of vanishing tensor interaction $(V_T=0)\,$ $\nu_L$
decouples from $\nu_R\,$. Setting also $V^{RR}=0$ and
$V^{LL}$ equal to the SM charged current and neutral current
interactions, we recover the SM case, with 
decoupled non-interacting right-handed states.


\section{Implications and Discussion}
\label{discussion}

Setting $g_V\,$=$\,-g'_A\,$=$\,g_A\,$=$\,-g'_V\,$=$\,1$ 
in~(\ref{VVA})  and $\Sigma^{SP}=\Sigma^{T}=0$ 
yields the SM result for the potential felt by  
an electron neutrino propagating in an 
electron background.  
Introducing a unit vector in the direction of the neutrino
momentum $\hat \bk=\bk/|\bk|$ 
and using the explicit expression
for the spin vector $s$~\cite{BGN} we 
reproduce the result~\cite{Nunokawa}
\be \label{SMresult} 
V_{\nu,\bar\nu}^{SM} =  \pm 
\sqrt{2} \, G_F \, n_e 
\left[1 
- \left<{(\bp + m_e\blambda)\cdot \hat {\bk} - \bp\cdot\blambda 
\over
E_e}
+
{(\hat \bk\cdot\bp) \, (\bp\cdot\blambda) 
\over E_e(m_e+E_e)}\right> \right] 
\ee
where the plus-sign refers to neutrinos and 
the minus-sign to antineutrinos. 

However, our main result is  that in the presence of a 
neutrino tensor
interaction with the background fermions, the neutrino can undergo
spin-flip. This effect is similar to the spin-precession induced by a
transverse magnetic field $B_\perp$ that couples to the neutrino
magnetic dipole moment $\mu_\nu\,$. In fact, if we substitute
in~(\ref{EOMLR}) the off-diagonal term $V^{LR}_\pm$ by $\mu_\nu
B_\perp$ we obtain the equation of motion for a neutrino that
propagates in a magnetic field.~\cite{magnetic} Of course 
the two scenarios originate from different physics, 
however, formally they can be treated in the same way.

In the simplest one generation case, a left-handed neutrino 
produced at $t=0$ and propagating for a time $t$ in a constant medium
will be converted into a right-handed neutrino with a probability
$
P_{\nu}^{LR}(t) = 
\sin^2 2\theta \, 
\sin^2 \left({\Delta E_{\nu} \, t / 2} \right) \,
$
where the effective mixing angle is given by
\be \label{mixing}
\sin^2 2\theta = {|2 V^{LR}_+|^2 \over (\Delta E_{\nu})^2}\,,
\ee
and   $\Delta E_{\nu}$ is given 
in~(\ref{DeltaE}). In the case of more than one
neutrino flavor, propagation in a medium with changing density can
lead to resonance effects in complete analogy to the magnetic field
induced resonant spin-flip.~\cite{magnetic} 

Now, let us discuss shortly the
results for different types of background matter.
First consider a medium where the average momentum of the background
fermions vanishes: $\vev{\bp} = 0$ ({\it e.g.}  
when the momentum distribution is isotropic). The 
tensor component determining the effective mixing~(\ref{mixing}) 
is given by  
\be \label{isotropic} 
|V^{LR}_+| = \sqrt{2} \, G_F \, n_f
\sqrt{|g_T|^2 + |g_T'|^2} \, \left< \lambda_\perp\> \left(
\sin^2\vartheta+ {m_f \over E_f}
\cos^2\vartheta\right)
\right>\,,
\ee
where $\vartheta$ denotes the angle between the momentum and the
transverse polarization of the background fermion, and
$\lambda_\perp=\sqrt{\lambda_1^2 + \lambda_2^2}\,$.  Note that
$|V^{LR}_+|$ vanishes if the neutrino propagates along the direction
of the average background polarization ($\lambda_\perp=0$). For a
non-relativistic background ($E_f \simeq m_f \gg p_i$) this yields 
$
|V^{LR}_+| = \sqrt{2}\,G_F \, n_f
\sqrt{|g_T|^2 + |g_T'|^2} \, \vev{\lambda_\perp} 
$
while in the ultra-relativistic limit we find 
$|V^{LR}_+| \propto \vev{\lambda_\perp\,\sin^2\vartheta}$. 
Finally, for a degenerate
background in the presence of a magnetic field, only the fermions in
the lowest Landau level contribute to the polarization, with the spin
oriented antiparallel to the momentum. In  
this case the background is not isotropic, and one obtains 
\be \label{landau}
|V^{LR}_+| = \sqrt{2}\,G_F \, n_f
\sqrt{|g_T|^2 + |g_T'|^2} \, 
\left<\lambda_\perp {m_f \over E_f}\right> \,,
\ee
which vanishes in the ultra-relativistic limit.

It is interesting to note that tensor interactions 
could result from  neutrino scalar
couplings after Fierz rearrangement. 
Consider the tree level Lagrangian
\be \label{phiint}
-{\cal L}_{\rm tree}=
\lambda_\phi \phi \, (\lbar{L_L} \, e_R) +
\lambda'_\phi \tilde\phi \, (\lbar{L_L} \, \nu_R) \, + {\rm h.c.}\,,
\ee
involving a right-handed neutrino singlet ($\nu_R$)  
and a doublet scalar field $\phi$ with mass $m_\phi$ and  
couplings $\lambda_\phi, \lambda'_\phi$ to the lepton fields.  
The resulting set of low energy  
effective interactions contains the following terms:  
\be \label{fourfermi}
{\cal H}_{\rm int}^\phi = 
-{\lambda'_\phi \lambda_\phi \over m_\phi^2}
\left[
\half(\lbar{\nu_R} \, \nu_L) \, (\lbar{e_R} \, e_L) +
\eighth
(\lbar{\nu_R} \, \sigma_{\mu \nu} \, \nu_L) \,
(\lbar{e_R} \, \sigma^{\mu \nu} \, e_L)
\right] \,,
\ee
implying $g_T \sim {\lambda'_\phi \lambda_\phi/ m_\phi^2}$.  
When different scalar fields mix,   
operators of this kind can be generated 
also in supersymmetric models without $R$-parity. 

Let us now address the issue whether the new tensor term could
be relevant for real physical systems, like the sun or a 
galactic supernova.
{}From eqs.~(\ref{isotropic})--(\ref{landau}) it follows
that, with respect to the SM vector potential,   
 the effective tensor potential is   
suppressed by a factor
\be \label{eps}
\epsilon \equiv \left| V^{LR}_+ \over V^{LL}_0 \right|
\lsim \sqrt{|g_T|^2 + |g_T'|^2} \, \langle \lambda_\perp \rangle\,.
\ee
New physics effects can be
relevant to neutrino oscillations only if 
$g_T, g_T'$ and $\langle \lambda_\perp \rangle$
are large enough to affect sizably the 
results obtained within the SM.
In particular $\epsilon$ should satisfy the 
lower limits:~\cite{BGN,BergmannKagan}
$
\epsilon_{sun} \gsim 10^{-2} $ and $ \epsilon_{SN} \gsim 10^{-4}. 
$ 
The excellent agreement between the SM predictions 
and various  experimental results,
suggests that $g_T$ and $g_T'$ are small, 
probably  not exceeding the few percent level.
However, the tiny values of the average
polarization is by far the most important suppression factor.  
In the solar 
interior, the magnetic field can be at most of the order of several
kG. This results in a very small electron polarization~\cite{BGN} 
$ \vev{\lambda_{e}} \simeq 10^{-8}$ and 
quite likely neutrino propagation in the sun cannot be affected by the
new tensor interaction.
{}For a proto-neutron star in the early cooling phase,
soon after the supernova explosion, 
the magnetic field strength can reach very large values.  
However, the temperature is also large, thus
suppressing the induced polarization.  
For a magnetic field $B\sim 10^{13}$\,G,
it was estimated~\cite{BGN} 
$ \vev{\lambda_{e}} \simeq 10^{-4}$ and  
for the nucleon polarization $\vev{\lambda_{p,n}} 
\simeq 10^{-5}$. 
Thus, if $B \lsim 10^{13}$\,G the propagation of 
supernova neutrinos would not be affected.
However, it was pointed out~\cite{BGN} 
that collision effects could  increase the
production of right-handed states and thus  
enhance the effects of the tensor interaction. 
Also, the value of the proto-neutron star magnetic field  
is poorly known. It has been proposed that at early
times it could be as large as 
$10^{16}\, $G.~\cite{strongB}  This would imply an
enhancement of the polarization of about three orders of magnitude,
opening the possibility of observing these effects. 

{}Finally, let us note that since the presence of right-handed
neutrinos implies in
general a non-vanishing magnetic moment, the effect of the tensor
interaction will be accompanied by similar effects due to 
the neutrino magnetic moment coupled to the strong magnetic field. 
Clearly, in this case 
both effects have to be taken into account simultaneously.


\end{document}